\documentclass[twocolumn,prl,superscriptaddress,showpacs,preprintnumbers]{revtex4}

\usepackage{hyperref,amsmath,graphicx}

\begin{document}


\title{Improved self-absorption correction for fluorescence measurements \\
of extended x-ray absorption fine-structure}

\author{C. H. Booth} \affiliation{Chemical Sciences Division, Lawrence
Berkeley National Laboratory, Berkeley, California 94720, USA}\email{chbooth@lbl.gov}

\author{F. Bridges} \affiliation{Physics Department, University of California, Santa Cruz, California 95064, USA}

\date{Submitted to the XAFS XII conference, June 9, 2003}

\preprint{LBNL-52864}

\begin{abstract}
Extended x-ray absorption fine-structure (EXAFS) data collected in the 
fluorescence mode are susceptible to an apparent amplitude reduction due to
the self-absorption of the fluorescing photon by the sample before it 
reaches a detector.  Previous treatments have made the simplifying assumption 
that the effect of the EXAFS on the correction term is negligible, and that the
samples are in the thick limit.  We present a nearly exact treatment
that can be applied for any sample thickness or concentration, and retains the
EXAFS oscillations in the correction term.
\end{abstract}



\maketitle


Under ideal circumstances, such as a very dilute sample, the photoelectric part 
of the x-ray absorption coefficient, $\mu$, is proportional to the number of 
fluorescence photons escaping the sample.  However, in extended x-ray 
absorption fine-structure spectroscopy (EXAFS), the mean absorption depth 
changes with the energy of the incident photon, 
$E$, which changes the probability that the fluorescence photon will be 
reabsorbed by the sample.  
This self-absorption causes a reduction in the measured EXAFS oscillations, 
$\chi_{exp}$, from the true $\chi$, and hence needs to be included in any subsequent 
analysis.  

Previous treatments \cite{Goulon82,Tan89,Troger92} to correct for the 
self-absorption effect 
account for the change in depth due to the absorption edge and due to the smooth
decrease in $\mu$ that follows, for instance, a Victoreen formula, and
have been shown to be quite effective in certain limits.  These 
treatments typically make two important assumptions.  First, the so-called
``thick limit'' is used to eliminate the dependence on the actual sample
thickness, limiting the applicability to thick, concentrated samples, such as
single crystals. One exception is the work of
Tan, Budnick and Heald \cite{Tan89}, which makes a number of other assumptions 
to estimate the correction to the amplitude reduction factor, $S_0^2$, and to the 
Debye-Waller factors, $\sigma^2$'s, rather than correcting the data in a
model-independent way.  A second assumption is that, in order to make the 
correction factor analytical, at 
one point in the calculation, the true absorption 
coefficients for the absorbing species and the whole sample are replaced with 
their average values; in other words, the modulating effect of $\chi$ on the
correction factor is taken as very small.
Below, we present a treatment that, with only one assumption that is nearly 
exact 
for all cases we have measured, corrects fluorescence EXAFS data directly in 
$k$-space for any concentration or thickness.  This correction is demonstrated 
for a copper foil that is about one absorption-length thick, and is therefore 
not in the thick limit.


\begin{figure}[t]
\center{\includegraphics[width=2.8in,trim=200 0 0 0,clip]{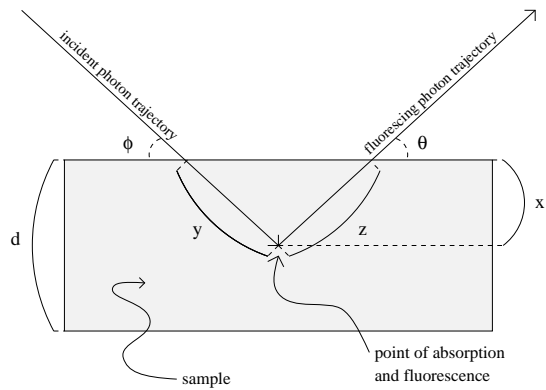}}
\caption{Geometry used in calculating self-absorption correction in EXAFS.}.
\label{geometry}
\end{figure}

Figure \ref{geometry} shows the geometry used in this calculation.
The fluorescence yield at the point of absorption is proportional to the x-ray
intensity $I$ at that point and the fluorescence efficiency.  The intensity $I$
at a depth $y$ is
\[
I = I_0 e^{-\mu(E) y}.
\]
The fluorescence photon then has to escape.  The fluorescence flux from this 
point in the sample is then
\[
I_f = I_0 e^{-\mu(E) y} e^{-\mu(E_f) z} \epsilon_a(E) \mu_a(E),
\]
where $\mu_a(E)$ is the absorption due to the absorbing atom, $\mu(E)$ is
the total absorption, $\epsilon_a(E)$ is the fluorescence efficiency per unit 
solid angle,
$E$ is the incident beam energy, $E_f$ is the energy of the fluorescing 
photon, and we're assuming that all the measured fluorescence is coming from
the desired process (eg. Cu $K_\alpha$, any other counts can be subtracted off).
This equation is only true at a particular $y$ and $z$, so we
must integrate
\[
dI_f = I_0 \epsilon_a \mu_a e^{-(\mu_T y + \mu_f z)} dy.
\]
Here the energy dependences are implicit and we've used $\mu_T = \mu(E)$ and
$\mu_f = \mu(E_f)$.  The variables $z$ and $y$ are dependent via 
$y \sin \phi = z \sin \theta = x$.  Changing variables, we obtain
\[ dI_f = I_0 \epsilon_a \mu_a \frac{1}{\sin \phi} 
e^{-(\frac{\mu_T}{\sin \phi} + \frac{\mu_f}{\sin \theta}) x} dx
\]
\begin{eqnarray}
I_f &=& I_0 \epsilon_a \mu_a \frac{1}{\sin \phi}
\int_0^d e^{-(\frac{\mu_T}{\sin \phi} + \frac{\mu_f}{\sin \theta}) x} dx
\nonumber\\
I_f &=& \frac{I_0 \epsilon_a \mu_a} {\mu_T + g \mu_f} 
\left [1 - e^{-(\frac{\mu_T}{\sin \phi} + \frac{\mu_f}{\sin \theta}) d } \right ],
\label{eq_totalfluor}
\end{eqnarray}
where $g \equiv \frac{\sin \phi}{\sin \theta}$.  
Eq. \ref{eq_totalfluor} describes the 
fluorescence in the direction given by $\theta$.  At
this point one should integrate over the detector's solid angle.
Ignoring this integral can affect the final obtained correction \cite{Brewe94}, 
especially for glancing-emergent angle experiments.  However,
for detector geometries where $\phi+\theta=90^\circ$, we find the maximum error 
in $g$ is on the order of $\sim1-2\%$ even for $\Delta\theta\approx 5^\circ$
at $\theta=80^\circ$.  For more severe geometries, the solid angle should be 
considered, but for the following, we ignore this correction.

In EXAFS measurements, we want
\[
\chi = \frac{\mu_a - \bar{\mu_a}}{\bar{\mu_a}},
\]
but what we actually obtain experimentally is
\[
\chi_{exp}= \frac{I_f - \bar{I_f}}{\bar{I_f}},
\]
where $\bar{I_f}$ is the spline function fit to the data to simulate the 
``embedded atom'' background 
fluorescence (roughly $I_f$ without the EXAFS oscillations).
Now make the following substitutions:
\begin{eqnarray*}
\mu_T &=& \bar{\mu_T} + \chi \bar{\mu_a}\\
\mu_a &=& (\chi + 1)\bar{\mu_a}\\
\mu_T - \bar{\mu_T} &=& \mu_a - \bar{\mu_a}.
\end{eqnarray*}
These equations and Eq. \ref{eq_totalfluor} are then plugged into $\chi_{exp}$:
\[
1 + \chi_{exp} = \frac{\mu_a (\bar{\mu_T} + g \mu_f) [1- e^{-(\frac{\mu_T}{\sin \phi} + \frac{\mu_f}{\sin \theta}) d }]}
{\bar{\mu_a}(\mu_T + g\mu_f)[1-e^{-(\frac{\bar{\mu_T}}{\sin \phi} + \frac{\mu_f}{\sin \theta}) d }]}.
\]
Dividing by $1+\chi$ and defining $\alpha \equiv \bar{\mu_T} + g \mu_f$,
we get:
\[
\frac{1+\chi_{exp}}{1+\chi}=\frac{[1- e^{-(\bar{\mu_T} + \chi\bar{\mu_a} 
+ g\mu_f)\frac{d}{\sin \phi}}] \alpha}
{(\alpha + \chi\bar{\mu_a})[1-e^{-\frac{\alpha d}{\sin \phi}}]}.
\]
Now $\chi_{exp}$ can be written in terms of the actual $\chi$:
\begin{equation}
\chi_{exp} = \left[ \frac{1-e^{-(\alpha + \chi \bar{\mu_a}) \frac{d}{\sin \phi}}}
{1-e^{-\frac{\alpha d}{\sin \phi}}} \right] \left[ \frac{\alpha (\chi + 1)}{\alpha + \chi
\bar{\mu_a}}\right] - 1.
\label{chi_exp}
\end{equation}
At this point in the calculation, the relation between $\chi$ and $\chi_{exp}$
is exact.  However, we need $\chi$ in terms of $\chi_{exp}$, and 
Eq.~\ref{chi_exp} is for $\chi_{exp}$ in terms of $\chi$.  In order to invert 
Eq.~\ref{chi_exp}, we make a simple approximation.  Assuming that
\[
\frac{\chi \bar{\mu_a} d}{\sin \phi} << 1
\]
we can say
\begin{equation}
1 - e^{-(\alpha + \chi\bar{\mu_a})\frac{d}{\sin \phi}} \sim 
1 - e^{-\frac{\alpha d}{\sin \phi}}(1-\frac{\chi\bar{\mu_a}d}{\sin \phi}).
\label{eq_approx}
\end{equation}
This approximation gets worse with large $\chi$ and $\bar{\mu_a}$.  It also
has a maximum for both $\phi$ and $d$, because of the 
$e^{-\frac{\alpha d}{\sin \phi}}$ term.  Plugging in some typical numbers from
the Cu $K$-edge of YBa$_2$Cu$_3$O$_7$ ($\phi = 10 ^\circ$, 
$\bar{\mu_T}=01.32 \mu$m$^{-1}, \mu_F=1.10 \mu$m$^{-1}, 
\bar{\mu_a} =1.0 \mu$m$^{-1}$ and $\chi = 0.5$) the maximum error is $\sim 2.7\%$ at a
thickness of $\sim 1.9 \mu$m.  Such a high value of $\chi$ does not actually
occur in YBCO.  Indeed, such a high $\chi$ is very rare.  In any case, various
combinations of the above parameters can conspire to produce errors above 1\%, 
so the approximation should be monitored when making the corrections outlined 
below. 

With the above approximation, and defining the following quantities:
\begin{eqnarray*}
\beta &=& \frac{\bar{\mu_a}d\alpha}{\sin \phi} 
e ^ {-\frac{\alpha d}{\sin \phi}}\\
\gamma &=& 1 - e^{-\frac{\alpha d}{\sin \phi}},
\end{eqnarray*}
Eq.~\ref{chi_exp} is reduced to a quadratic equation in $\chi$ and
we can finally write the full correction formula:
\begin{widetext}
\begin{equation}
\chi = \frac{ -[\gamma (\alpha - \bar{\mu_a}(\chi_{exp}+1)) + \beta] +
\sqrt{ [\gamma (\alpha - \bar{\mu_a}(\chi_{exp}+1)) + \beta]^2 + 
4\alpha\beta\gamma\chi_{exp} } }
{2\beta},
\label{FY_corr}
\end{equation}
\end{widetext}
where the sign of the square root was determined by taking the thick or
thin limits.
In the thick limit ($d \rightarrow \infty$), Eq.~\ref{FY_corr} gives:
\[
\chi=\frac{\chi_{exp}}{1-\frac{\bar{\mu_a}}{\alpha}\chi_{exp} - 
\frac{\bar{\mu_a}}{\alpha}},
\]
which is the same as calculated in Ref. \cite{Troger92} without 
the $\chi_{exp}$
term in the denominator.  In the thin limit, it can be shown that 
Eq.~\ref{FY_corr} reduces to $\chi = \chi_{exp}$, as expected.

\begin{figure}[t]
\includegraphics[width=3.4in,trim=0 10 0 0,clip]{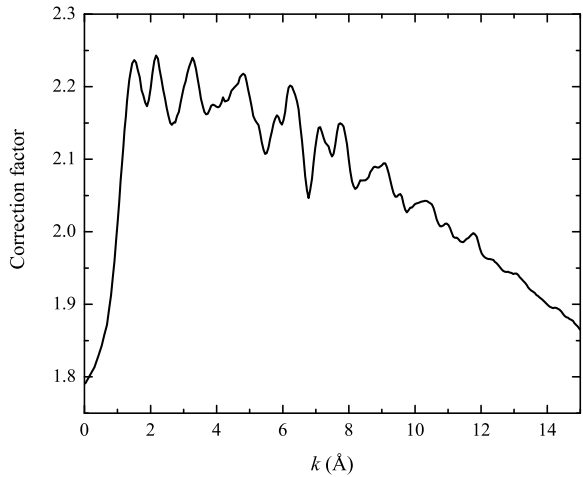}
\caption{Correction term $\chi/\chi_{exp}$
given by Eq.~\protect\ref{FY_corr} for Cu $K$-edge 
absorption data from a 4.6 $\mu$m-thick copper foil at $\phi=49.4^\circ$.
}
\label{sab_xtal}
\end{figure}

\begin{figure}[t]
\includegraphics[width=3.4in,trim=0 10 0 0,clip]{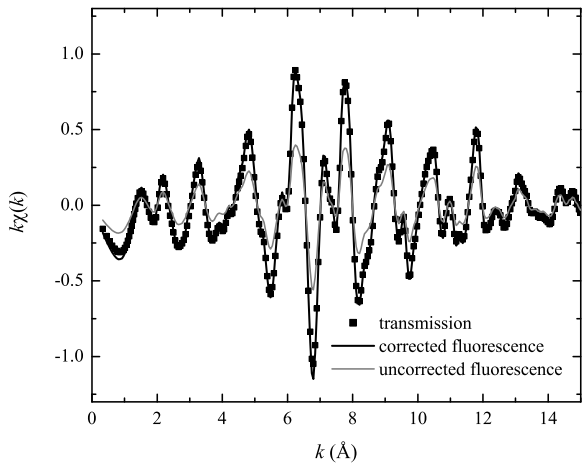}
\caption{Corrected EXAFS data in $k$-space for the copper foil data, compared 
to transmission data and uncorrected fluorescence data.  Note that the 
corrected data are difficult to discern on top of the transmission data.
}
\label{kspace}
\end{figure}


We performed an experiment on a copper foil to demonstrate the correction.
Cu $K$-edge data were 
collected both in the transmission mode and in the 
fluorescence mode
using a 32-element Canberra germanium detector on beam line 11-2 at 
the Stanford Synchrotron Radiation Laboratory (SSRL).  The transmission data 
were checked for pinhole effects (found to be negligible) and the fluorescence
data were corrected for dead time.  The sample thickness was estimated to be 
4.6 $\mu$m from 
the absorption step at the edge, and was oriented such that 
$\phi=49.4\pm0.5^\circ$. The thickness is about 25\% of the estimated 
thick-limit thickness.  The data were reduced to $k$-space using the
RSXAP analysis program REDUCE \cite{Hayes82,Li95b,RSXAP}, which 
incorporates these corrections.
Figure \ref{sab_xtal} shows the correction factor ($\chi/\chi_{exp}$) for these data.
The error in the approximation in Eq. \ref{eq_approx} exceeds 1\% only
below $\sim 1 $\AA$^{-1}$. The total
correction in the thick limit is much larger (about 3 times the displayed
correction).
As shown in Fig. \ref{kspace}, the corrected
fluorescence data in $k$-space are remarkably similar to the transmission data,
despite the large magnitude of the correction.

Although only a copper foil is reported as an example, we have successfully 
applied this correction to a wide range of oxides and intermetallics, including 
single crystals and thin films \cite{Booth96,Ren98,Cao00,Mannella03,Booth03}.  The 
ability to correct for intermediate film
thicknesses is, in fact, crucial for studying films thinner than 
$\sim 20 \mu$m thick.

In summary, we have provided an improved self-absorption correction for EXAFS 
data that operates at any sample thickness or concentration.  Our example of a
pure copper foil demonstrates both the accuracy of the correction and that,
for concentrated samples, the correction can be surprisingly large.  Moreover,
for well-ordered materials, $\chi$ can have a surprisingly large effect. 

\acknowledgments

This work was partially supported by the Director, Office of Science, Office of Basic Energy Sciences (OBES), Chemical Sciences, Geosciences and Biosciences Division, U. S. Department of Energy (DOE) under Contract No. AC03-76SF00098.  
EXAFS data were collected at SSRL, a national user facility operated by 
Stanford University on behalf of the DOE/OBES.

\bibliographystyle{apsrev}
\bibliography{/home/hahn/chbooth/papers/bib/bibli}

\begin{thebibliography}{12}
\expandafter\ifx\csname natexlab\endcsname\relax\def\natexlab#1{#1}\fi
\expandafter\ifx\csname bibnamefont\endcsname\relax
  \def\bibnamefont#1{#1}\fi
\expandafter\ifx\csname bibfnamefont\endcsname\relax
  \def\bibfnamefont#1{#1}\fi
\expandafter\ifx\csname citenamefont\endcsname\relax
  \def\citenamefont#1{#1}\fi
\expandafter\ifx\csname url\endcsname\relax
  \def\url#1{\texttt{#1}}\fi
\expandafter\ifx\csname urlprefix\endcsname\relax\def\urlprefix{URL }\fi
\providecommand{\bibinfo}[2]{#2}
\providecommand{\eprint}[2][]{\url{#2}}

\bibitem[{\citenamefont{Goulon et~al.}(1982)\citenamefont{Goulon,
  {Goulon-Ginet}, Cortes, and Dubois}}]{Goulon82}
\bibinfo{author}{\bibfnamefont{J.}~\bibnamefont{Goulon}},
  \bibinfo{author}{\bibfnamefont{C.}~\bibnamefont{{Goulon-Ginet}}},
  \bibinfo{author}{\bibfnamefont{R.}~\bibnamefont{Cortes}}, \bibnamefont{and}
  \bibinfo{author}{\bibfnamefont{J.~M.} \bibnamefont{Dubois}},
  \bibinfo{journal}{J. Physique} \textbf{\bibinfo{volume}{43}},
  \bibinfo{pages}{539} (\bibinfo{year}{1982}).

\bibitem[{\citenamefont{Tan et~al.}(1989)\citenamefont{Tan, Budnick, and
  Heald}}]{Tan89}
\bibinfo{author}{\bibfnamefont{Z.}~\bibnamefont{Tan}},
  \bibinfo{author}{\bibfnamefont{J.~I.} \bibnamefont{Budnick}},
  \bibnamefont{and} \bibinfo{author}{\bibfnamefont{S.~M.} \bibnamefont{Heald}},
  \bibinfo{journal}{Rev. Sci. Instrum.} \textbf{\bibinfo{volume}{60}},
  \bibinfo{pages}{1021} (\bibinfo{year}{1989}).

\bibitem[{\citenamefont{Tr{\"{o}}ger et~al.}(1992)\citenamefont{Tr{\"{o}}ger,
  Arvanitis, Baberschke, Michaelis, Grimm, and Zschech}}]{Troger92}
\bibinfo{author}{\bibfnamefont{L.}~\bibnamefont{Tr{\"{o}}ger}},
  \bibinfo{author}{\bibfnamefont{D.}~\bibnamefont{Arvanitis}},
  \bibinfo{author}{\bibfnamefont{K.}~\bibnamefont{Baberschke}},
  \bibinfo{author}{\bibfnamefont{H.}~\bibnamefont{Michaelis}},
  \bibinfo{author}{\bibfnamefont{U.}~\bibnamefont{Grimm}}, \bibnamefont{and}
  \bibinfo{author}{\bibfnamefont{E.}~\bibnamefont{Zschech}},
  \bibinfo{journal}{Phys. Rev. B} \textbf{\bibinfo{volume}{46}},
  \bibinfo{pages}{3283} (\bibinfo{year}{1992}).

\bibitem[{\citenamefont{Brewe et~al.}(1994)\citenamefont{Brewe, Pease, , and
  Budnick}}]{Brewe94}
\bibinfo{author}{\bibfnamefont{D.~L.} \bibnamefont{Brewe}},
  \bibinfo{author}{\bibfnamefont{D.~M.} \bibnamefont{Pease}}, ,
  \bibnamefont{and} \bibinfo{author}{\bibfnamefont{J.~I.}
  \bibnamefont{Budnick}}, \bibinfo{journal}{Phys. Rev. B}
  \textbf{\bibinfo{volume}{50}}, \bibinfo{pages}{9025} (\bibinfo{year}{1994}).

\bibitem[{\citenamefont{Hayes and Boyce}(1982)}]{Hayes82}
\bibinfo{author}{\bibfnamefont{T.~M.} \bibnamefont{Hayes}} \bibnamefont{and}
  \bibinfo{author}{\bibfnamefont{J.~B.} \bibnamefont{Boyce}}, in
  \emph{\bibinfo{booktitle}{Solid State Physics}}, edited by
  \bibinfo{editor}{\bibfnamefont{H.}~\bibnamefont{Ehrenreich}},
  \bibinfo{editor}{\bibfnamefont{F.}~\bibnamefont{Seitz}}, \bibnamefont{and}
  \bibinfo{editor}{\bibfnamefont{D.}~\bibnamefont{Turnbull}}
  (\bibinfo{publisher}{Academic}, \bibinfo{address}{New York},
  \bibinfo{year}{1982}), vol.~\bibinfo{volume}{37}, p. \bibinfo{pages}{173}.

\bibitem[{\citenamefont{Li et~al.}(1995)\citenamefont{Li, Bridges, and
  Booth}}]{Li95b}
\bibinfo{author}{\bibfnamefont{G.~G.} \bibnamefont{Li}},
  \bibinfo{author}{\bibfnamefont{F.}~\bibnamefont{Bridges}}, \bibnamefont{and}
  \bibinfo{author}{\bibfnamefont{C.~H.} \bibnamefont{Booth}},
  \bibinfo{journal}{Phys. Rev. B} \textbf{\bibinfo{volume}{52}},
  \bibinfo{pages}{6332} (\bibinfo{year}{1995}).

\bibitem[{RSX()}]{RSXAP}
\eprint{http://lise.lbl.gov/RSXAP/}.

\bibitem[{\citenamefont{Booth et~al.}(1996)\citenamefont{Booth, Bridges, Boyce,
  Claeson, Lairson, Liang, and Bonn}}]{Booth96}
\bibinfo{author}{\bibfnamefont{C.~H.} \bibnamefont{Booth}},
  \bibinfo{author}{\bibfnamefont{F.}~\bibnamefont{Bridges}},
  \bibinfo{author}{\bibfnamefont{J.}~\bibnamefont{Boyce}},
  \bibinfo{author}{\bibfnamefont{T.}~\bibnamefont{Claeson}},
  \bibinfo{author}{\bibfnamefont{B.~M.} \bibnamefont{Lairson}},
  \bibinfo{author}{\bibfnamefont{R.}~\bibnamefont{Liang}}, \bibnamefont{and}
  \bibinfo{author}{\bibfnamefont{D.~A.} \bibnamefont{Bonn}},
  \bibinfo{journal}{Phys. Rev. B} \textbf{\bibinfo{volume}{54}},
  \bibinfo{pages}{9542} (\bibinfo{year}{1996}).

\bibitem[{\citenamefont{Ren et~al.}(1998)\citenamefont{Ren, Rose, Williams,
  Booth, Shuh, Allen, Bucher, and Edelstein}}]{Ren98}
\bibinfo{author}{\bibfnamefont{J.~Z.} \bibnamefont{Ren}},
  \bibinfo{author}{\bibfnamefont{G.~A.} \bibnamefont{Rose}},
  \bibinfo{author}{\bibfnamefont{R.~S.} \bibnamefont{Williams}},
  \bibinfo{author}{\bibfnamefont{C.~H.} \bibnamefont{Booth}},
  \bibinfo{author}{\bibfnamefont{D.~K.} \bibnamefont{Shuh}},
  \bibinfo{author}{\bibfnamefont{P.~G.} \bibnamefont{Allen}},
  \bibinfo{author}{\bibfnamefont{J.~J.} \bibnamefont{Bucher}},
  \bibnamefont{and} \bibinfo{author}{\bibfnamefont{N.~M.}
  \bibnamefont{Edelstein}}, \bibinfo{journal}{J. Appl. Phys.}
  \textbf{\bibinfo{volume}{83}}, \bibinfo{pages}{7613} (\bibinfo{year}{1998}).

\bibitem[{\citenamefont{Mannella et~al.}(2003)\citenamefont{Mannella,
  Rosenhahn, Booth, Marchesini, Mun, {S.-H. Yang}, Ibrahim, Tomioka, and
  Fadley}}]{Mannella03}
\bibinfo{author}{\bibfnamefont{N.}~\bibnamefont{Mannella}},
  \bibinfo{author}{\bibfnamefont{A.}~\bibnamefont{Rosenhahn}},
  \bibinfo{author}{\bibfnamefont{C.~H.} \bibnamefont{Booth}},
  \bibinfo{author}{\bibfnamefont{S.}~\bibnamefont{Marchesini}},
  \bibinfo{author}{\bibfnamefont{B.~S.} \bibnamefont{Mun}},
  \bibinfo{author}{\bibnamefont{{S.-H. Yang}}},
  \bibinfo{author}{\bibfnamefont{K.}~\bibnamefont{Ibrahim}},
  \bibinfo{author}{\bibfnamefont{Y.}~\bibnamefont{Tomioka}}, \bibnamefont{and}
  \bibinfo{author}{\bibfnamefont{C.~S.} \bibnamefont{Fadley}}
  (\bibinfo{year}{2003}), \bibinfo{note}{preprint}.

\bibitem[{\citenamefont{Booth et~al.}()\citenamefont{Booth, Shlyk, Nenkov,
  Huber, and {L. E. De Long}}}]{Booth03}
\bibinfo{author}{\bibfnamefont{C.~H.} \bibnamefont{Booth}},
  \bibinfo{author}{\bibfnamefont{L.}~\bibnamefont{Shlyk}},
  \bibinfo{author}{\bibfnamefont{K.}~\bibnamefont{Nenkov}},
  \bibinfo{author}{\bibfnamefont{J.~G.} \bibnamefont{Huber}}, \bibnamefont{and}
  \bibinfo{author}{\bibnamefont{{L. E. De Long}}}, \bibinfo{note}{submitted}.

\bibitem[{\citenamefont{Cao et~al.}(2000)\citenamefont{Cao, Bridges, Worledge,
  Booth, and Geballe}}]{Cao00}
\bibinfo{author}{\bibfnamefont{D.}~\bibnamefont{Cao}},
  \bibinfo{author}{\bibfnamefont{F.}~\bibnamefont{Bridges}},
  \bibinfo{author}{\bibfnamefont{D.~C.} \bibnamefont{Worledge}},
  \bibinfo{author}{\bibfnamefont{C.~H.} \bibnamefont{Booth}}, \bibnamefont{and}
  \bibinfo{author}{\bibfnamefont{T.}~\bibnamefont{Geballe}},
  \bibinfo{journal}{Phys. Rev. B} \textbf{\bibinfo{volume}{61}},
  \bibinfo{pages}{11373} (\bibinfo{year}{2000}).

\end{thebibliography}

\end{document}